\documentclass[acus]{JAC2000}

\usepackage{graphicx,epsfig,psfig}

\newcommand{\bi}{\begin{itemize}}
\newcommand{\ei}{\end{itemize}}

\newcommand{\rt}{\rightarrow}

\def\Journal#1#2#3#4{{#1} {\bf #2}, #3 (#4)}

\def\NIM{\em Nucl. Instrum. Methods}

\def\PLB{{\em Phys. Lett.}  B}
\def\PRL{\em Phys. Rev. Lett.}

\begin{document}
\title{\flushright{M10}\\[15pt] \centering
\boldmath  RESULTS AND FUTURE PLANS FROM BES}

\author{Zhengguo ZHAO, IHEP of CAS, Beijing, China}

\maketitle

\begin{abstract}

The values of $R = \sigma(e^+e^-\rightarrow\mbox{hadrons})/
\sigma(e^+e^-\rightarrow\mu^+\mu^-)$ for 85
center-of-mass energies between 2 and 5 GeV are reported. 
Preliminary results using partial wave analysis for $J/\psi$ 
decays to $\gamma \pi^+ \pi^-$, $\gamma K^+K^-$, $\phi \pi^+\pi^-$ 
and $\phi K^+K^-$ are presented. The BESIII/BEPCII, a project for the 
future of BES, is introduced.

\end{abstract}

\section{Present status of BESII/BEPC}

The Beijing Spectrometer (BES) at Beijing Electron Positron
Collider (BEPC), so far the only $e^+e^-$ machine operating in the
center-of-mass (cm) energy from 2-5 GeV and directly producing
charmonium and charmed mesons, has been running for 12 years.    
Both BEPC and BES were upgraded from 1995 to 1997, and the upgraded BES is
called BESII~\cite{bes2}. Currently the BES experiment is benefiting from the
upgraded machine and detector, though some of the detector and machine 
parts are suffering aging problems seriously. Table~\ref{tab:bespar} 
lists some major parameters of the detector performance. 

\begin{table*}[htp]
\begin{center}
\caption {Comparison of the major parameters with BESI and BESII}
\vspace{0.2cm}
\begin{tabular}{cccc}
\hline
 System & Parameter           & BESI          & BESII          \\\hline
 VC     & $\sigma_{xy}$($\mu$)& 250 (CDC)     & 90                 \\
        & planes              & 4             & 12                 \\
 MDC    & $\sigma_{xy}$($\mu$) & 200-250       & 190-210            \\
        & $\Delta p/p$        & 1.76$\sqrt{(1+p^2)}$ & 1.78$\sqrt{(1+p^2)}$
\\
 TOF    & $\sigma_t$(ps)      & 370           & 180                \\
        & $L_{attenuation}$(m)& 1-1.2         & 3.5-5.5            \\
 BSC    & $\Delta E/E(\%)$    & 25            & 21                \\
 ESC    & $\Delta E/E(\%)$    & 23            & 22                \\
 DAQ    & dead time/event(ms) & 20            & 10                 \\\hline
\end {tabular}
\end{center}
\label{tab:bespar}
\end{table*}

\begin{figure*}[!htb]
\begin{center}
\includegraphics*[width=120mm]{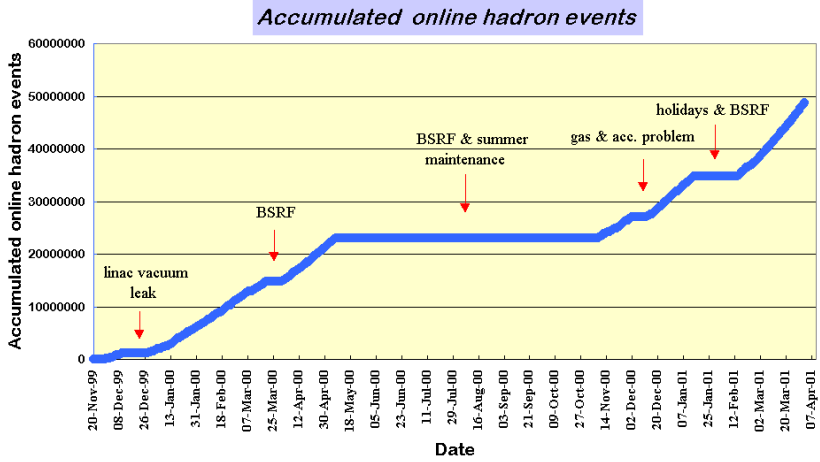}
\caption{Integraded hadronic events accumulated with BESII since Nov. 1999.}
\label{fig:jpsi}
\end{center}
\end{figure*}

Figure~\ref{fig:jpsi} shows the history of the integrated hadronic events 
accumulated with BESII since Nov. 1999 at the $J/\psi$ resonance. 
With this $50 \times 10^6$ $J/\psi$ event sample, which is about 6 times 
as high as the previous world's largest $J/\psi$ sample, BES can systematically 
study $J/\psi$ decays to  excited baryonic states and search for glueballs 
and hybrids.       

\section{Values of $R$ in 2-5 GeV}

In precision tests of the Standard Model (SM)~\cite{rlowe},
the quantities
$\alpha(M_Z^2)$, the QED running coupling constant evaluated at
the $Z$ pole, and $a_\mu = (g-2)/2$, the anomalous magnetic moment
of the muon, are of fundamental importance.

The dominant uncertainties in both $\alpha(M^2_{Z})$ and $a_{\mu}^{SM}$ are
due to the effects of hadronic vacuum polarization, which cannot be
reliably calculated in the low energy region.  Instead, with the
application of dispersion relations,
experimentally measured $R$ values are used to determine the vacuum
polarization, where $R$ is the lowest order cross section
for $e^+e^-\rightarrow\gamma^*\rightarrow \mbox{hadrons}$
in units of the lowest-order QED cross section for
$e^+e^- \rightarrow \mu^+\mu^-$, namely
$R=\sigma(e^+e^- \rightarrow \mbox{hadrons})/\sigma(e^+e^-\rightarrow
\mu^+\mu^-)$, where
$\sigma (e^+e^- \rightarrow \mu^+\mu^-) = \sigma^0_{\mu \mu}=
4\pi \alpha^2(0) / 3s$.
 
\begin{figure}[!htb]
\epsfysize=3.5in
\centerline{\epsfbox{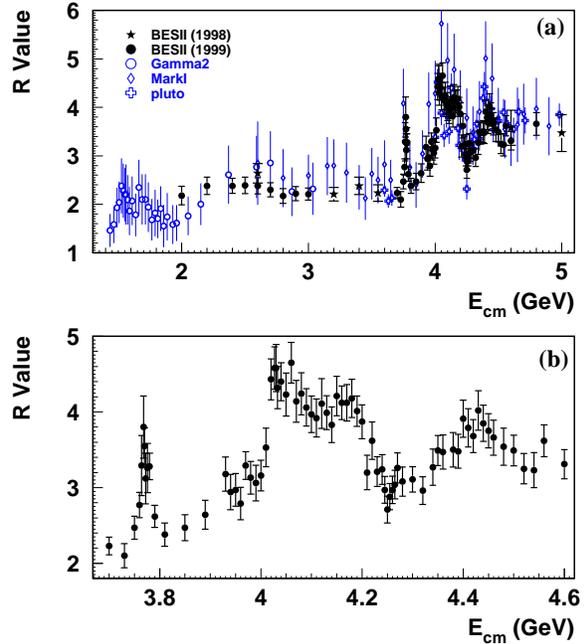}}
\caption{(a) A compilation of measurements of $R$ in the cm
energy range from 1.4 to 5 GeV. (b) $R$ values from this experiment
in the resonance region between 3.75 and 4.6 GeV.}
\label{fig:besr}
\end{figure}

The values of $R$ measured with BESII in 2-5 GeV are displayed in
Fig.~\ref{fig:besr}, together with BESII values from ref.~\cite{besr_1} 
and those measured by MarkI, $\gamma\gamma 2$, and
Pluto~\cite{mark1,gamma2,pluto}. 
The $R$ values from BESII have an average uncertainty of
about 6.6\%, which represents a factor of two to three improvement 
in precision in the 2 to 5 GeV
energy region.  Of this error, 4.3 \% is common to all points.
These improved measurements have a significant impact on the
global fit to the electroweak data and
the determination of the SM prediction for the mass
of the Higgs particle \cite{bolekpl}.   In addition, they
are expected to provide an
improvement in the precision of the calculated value of
$a_{\mu}^{SM}$~\cite{ichep2k,martin}, and test the QCD sum rules
down to 2 GeV~\cite{dave,kuehn}.

As pointed out by F.A. Harris~\cite{fred}, even after using the new 
BESII $R$ value, the error on $\alpha(M^2_{Z})$ is still dominated 
by the energy region from 1 to 5
GeV. Roughly 50\% of the error in $\alpha(M^2_{Z})$ is from 1-3 GeV.
Therefore, PEPN will make a great contribution to reduce this crucial
uncertainty. 

Using $R$ data below 3 GeV, we are also measuring the the cross section of 
$e^+e^- \rightarrow p \bar{p}$, the $\pi$ form factor, and testing pQCD by
studying hadronic event shape, energy dependence of inclusive spectra,
and some of the exclusive spectra (e.g. $e^+e^- \rightarrow 
\pi^+\pi^- X, K^+K^-X$) in $e^+e^-$ annihilation.   

\section{Resonance parameters of $\psi(2S)$ and $\psi(3770)$}

\begin{figure}[!htb]
\epsfysize=3.0in
\centerline{\epsfbox{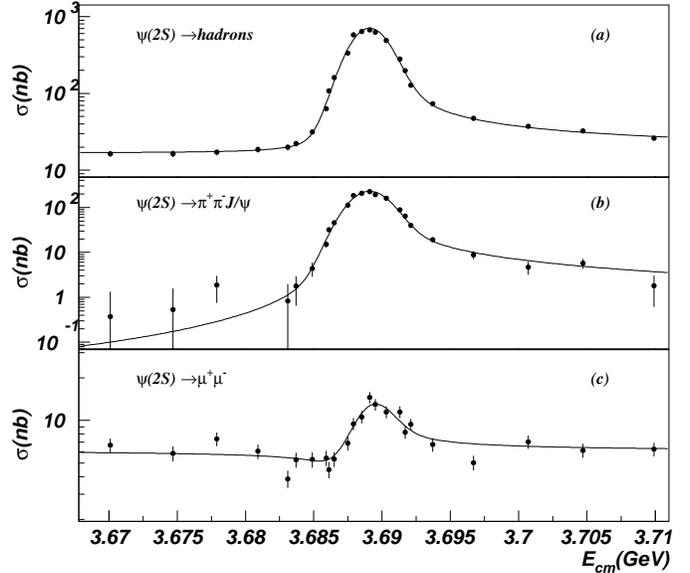}}
\caption{Measured cross section of $\psi(2S)\rightarrow hadron,
\pi^+\pi^-J/\psi$ and $\mu^+\mu^-$.}
\label{fig:psipscan}                                           
\end{figure}

To improve the measurement of the $\psi(2S)$ resonance parameters and
leptonic decay branching fraction, we did a detailed scan in the 3.67-3.71 
GeV region during the $R$ scan. Figure~\ref{fig:psipscan} shows 
the production cross section of 
$\psi(2S) \rightarrow {\rm hadrons},~\pi^+\pi^- J/\psi$, and $\mu^+\mu^-$.

\begin{figure}[!htb]
\epsfysize=4.0 in
\centerline{\epsfbox{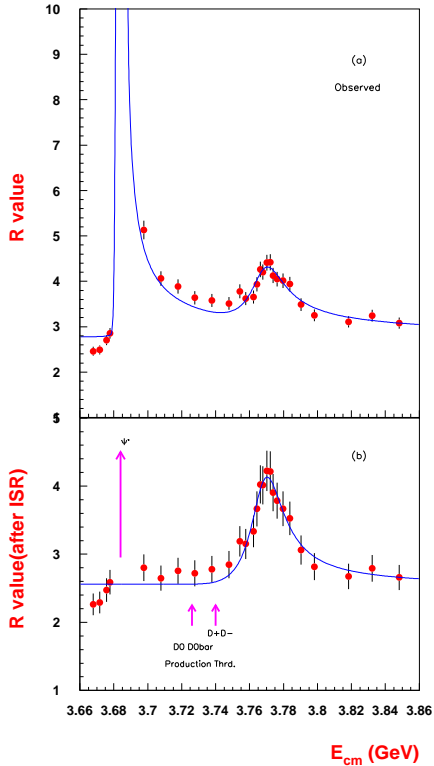}}
\caption{Preliminary $R$ values between 3.66 and 3.83 GeV from the detailed
scan done in the Spring of 2001 with BESII.}
\label{fig:psipp_scan}
\end{figure}

A detailed scan was also done this year over the $J/\psi$, $\psi(2S)$ and 
$\psi(3770)$ to improve the measurement of the resonance parameters 
and total production cross section of $\psi(3770)$. The $J/\psi$ and 
$\psi(2S)$ resonances are used for mass calibration. 
Figure~\ref{fig:psipp_scan} plots the preliminary $R$-values in the 3.66-3.83
GeV energy range.   

\section{$J/\psi$ DECAYS}

\begin{figure*}[htbp]
\centerline{
\epsfig{file=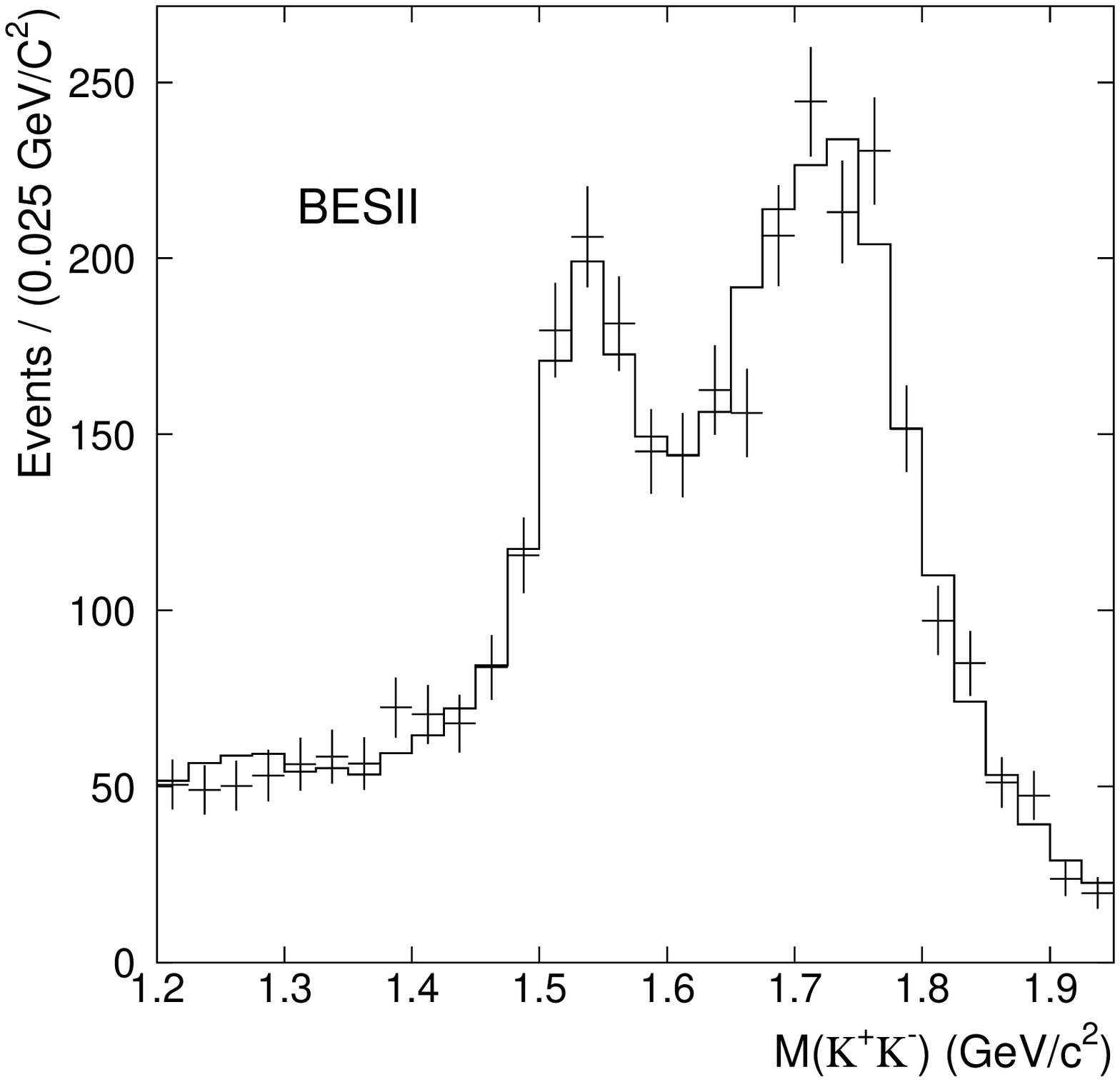,height=5.5cm,width=6.cm}
\epsfig{file=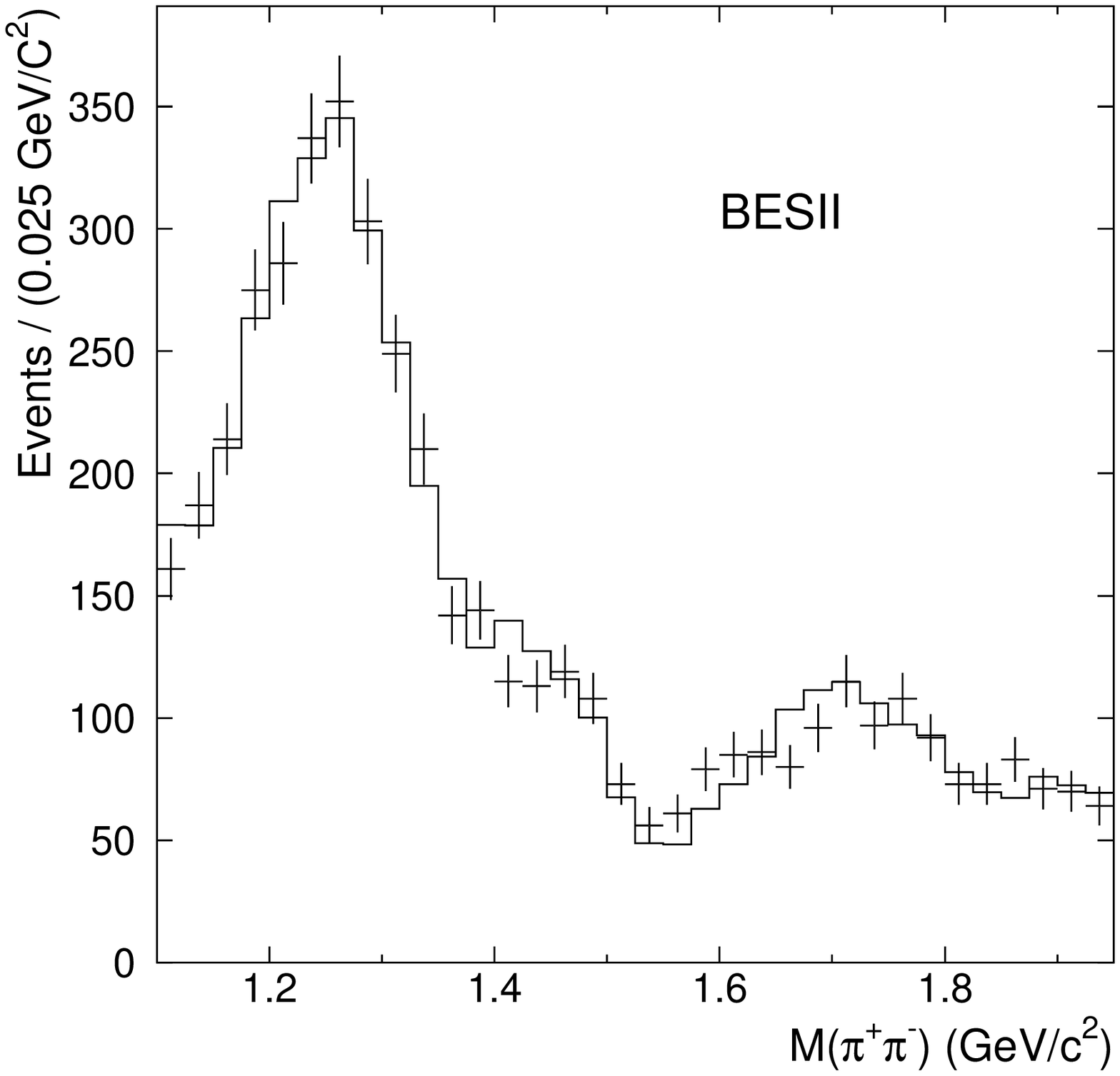,height=5.5cm,width=6.cm}}
\caption{Invariant mass spectra of $K^+ K^-$ for $J/\psi \to \gamma K^+ K^-$
(left) and $\pi^+ \pi^-$ for $J/\psi \to \gamma \pi^+ \pi^-$ (right). The
crosses are data and histograms the fits.}
\label{fig:jpsi1}
\end{figure*}

The $f_0(1710)$, first observed by the Crystal Ball Collaboration
in $J/\psi \to \gamma \eta \eta$, has been considered as the
possible lightest $0^{++}$ glueball candidate because of its large
production rate in gluon rich processes, such as $J/\psi$ radiative
decays, $p p$ central production, etc., and because of the
lattice QCD calculation of the lightest $0^{++}$ mass.
However, the spin-parity of $f_0(1710)$ is still not clear in
different channels after many years' efforts. Based on BESII
$24 \times 10^6$ $J/\psi$ data, the partial
wave analysis (PWA) is applied to the 1.7 GeV mass region in
$J/\psi$ radiative decays to $K^+ K^-$ and $\pi^+ \pi^-$.
In $K^+ K^-$ mass spectrum, the PWA analysis shows
a strong $2^{++}$ component, which is consistent with the well known $f_2'(1525)$
with the mass and width being in good agreement with PDG values,
and a dominant $0^{++}$ component in 1.7 GeV mass region. In the $J/\psi \to
\gamma \pi^+ \pi^-$ channel, there are two $0^{++}$ components, 
one located at around 1.5 GeV and another around 1.71 GeV, in addition to the
$f_2(1270)$. Fig.~\ref{fig:jpsi1} shows the $K^+ K^-$ and 
$\pi^+ \pi^-$ invariant mass spectra.
%

\begin{figure*}[htbp]
\centerline{
\epsfig{file=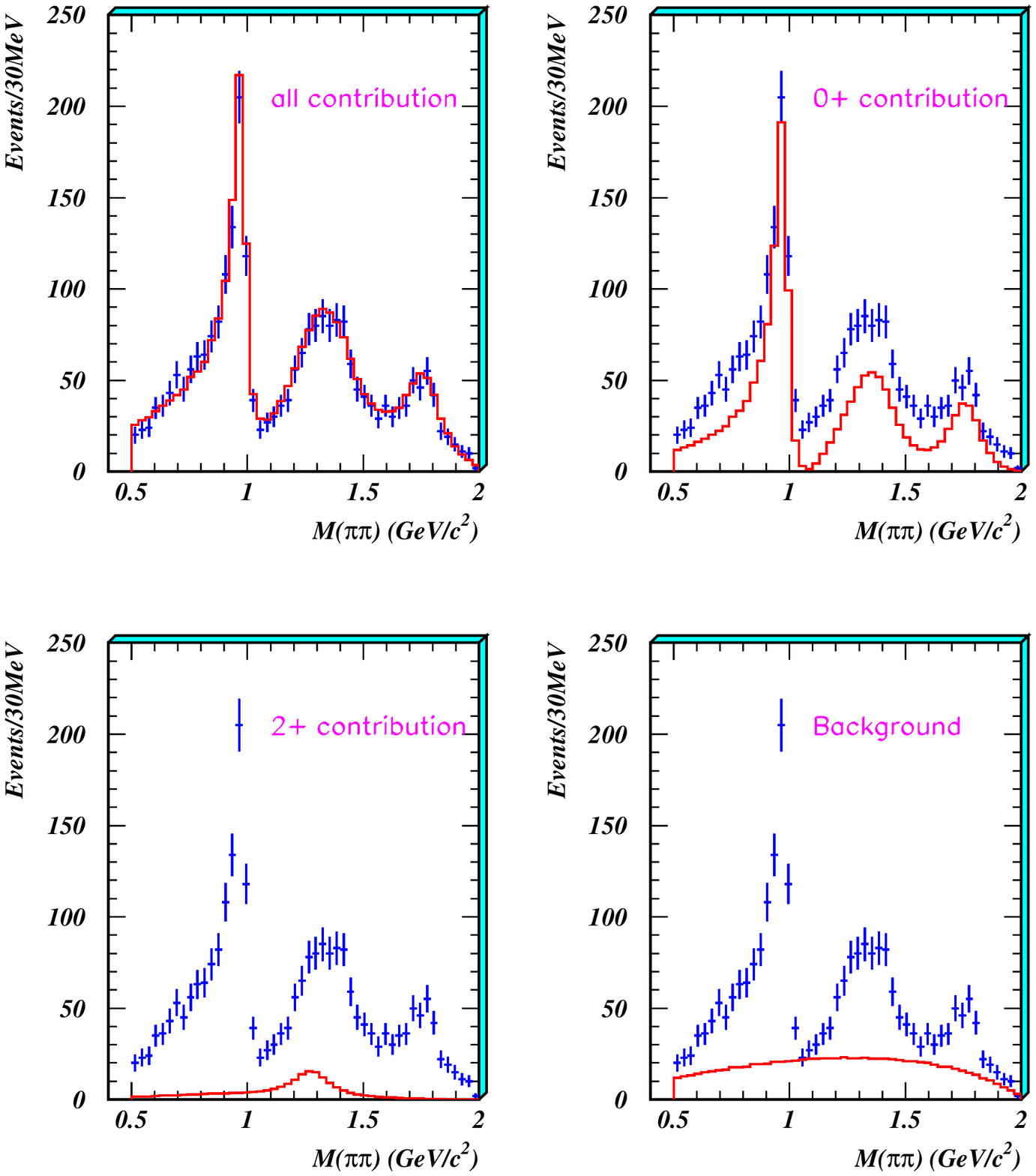,height=6.5cm,width=8.2cm}
\epsfig{file=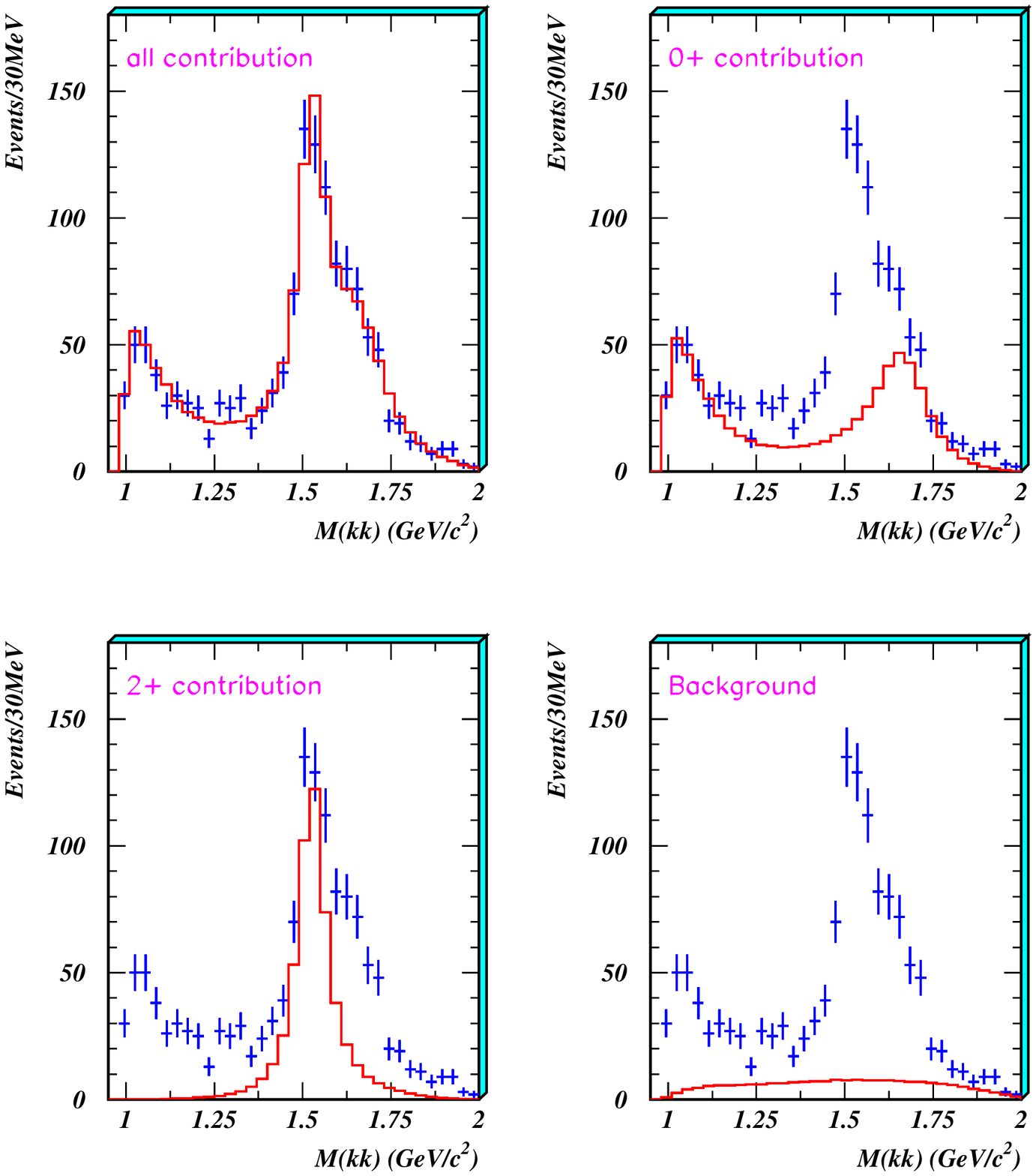,height=6.5cm,width=8.2cm}}
\caption{The mass projections of $\pi^+ \pi^-$ for $J/\psi \to \phi
\pi^+ \pi^-$ (left 4 plots) and $K^+ K^-$ for $J/\psi \to \phi K^+ K^-$
(right 4 plots). The crosses are data and histograms the fits.}
\label{fig:jpsi2}
\end{figure*}

With BESII $24 \times 10^6$ $J/\psi$ data, we performed a PWA analysis
of $J/\psi \to \phi \pi^+ \pi^-$ and $\phi K^+ K^-$. Fig.~\ref{fig:jpsi2} 
shows
the contribution of each component from the fit to $\phi \pi^+ \pi^-$
and $\phi K^+ K^-$. Three $0^{++}$ resonances, located at 980 MeV,
1370 MeV and 1770 MeV and one $2^{++}$ at 1270 MeV are observed
in the $\pi^+ \pi^-$ invariant mass spectrum recoiling against the $\phi$.
In $J/\psi \to \phi K^+ K^-$, the dominant resonance is $f_2'(1525)$ in
addition to the tail of $f_0(980)$. A further study on the shoulder of
$f_2'(1525)$ is needed.

MarkIII, DM2, and BES all found a broad structure in the lower
mass region of $\pi^+ \pi^-$ in $J/\psi \to \omega \pi^+ \pi^-$. Based on
$7.8\times 10^6$ BESI $J/\psi$ data, BES analyzed this channel and
found a $0^{++}$ wave to be required there.

\section{Future Plans}

The short term plan for the next 2-3 years is to continue running BESII, 
most likely to accumulate data at $\psi(2S)$ and $\psi(3770)$. The final 
decision will be made at the coming BES annual meeting.  
The long term plan for the future is the so called BESIII/BEPCII, a project 
to significantly upgrade both machine and detector. There are two 
options for the machine, i.e. a single ring with multi-bunch trains, or a 
double ring machine using the same machine tunnel. A single ring option is 
expected to have a luminosity about $3 \times 10^{32} cm^{-2}\cdot s^{-1}$ 
at the $J/\psi$ resonance. However, the luminosity provided by a double 
ring machine is expected to be at the level of $10^{33}
cm^{-2}\cdot s^{-1}$. The upgraded machine is called BEPCII.  

To match the BEPCII, BESII must be significantly upgraded. Particularly 
the capability of particle identification and photon detection must be
greatly enhanced.  Currently the detector has not been well defined. 
One possibility is to make use of the L3 BGO crystals as an 
electromagnetic calorimeter (EMCAL) with a tracking chamber inside and a
time-of-flight counter outside it. The muon identifier, which is out
side the superconducting coil will also be rebuilt.   
Another possibility is to build a KLOE type EMCAL
, which may provide an energy resolution of about $8\% \sqrt{E}$. 
In addition, new trigger, DAQ and electronics systems should be built to
adapt to the new beam characteristics with high luminosity. 

So far, \$40 M has been endorsed by the Chinese central government for
BESIII/BEPCII, and another \$20 M is under discussion. An
international review meeting was held in early April of this year for the
feasibility study. The double ring machine option is strongly favored by the
review committee. A proposal for the BESIII/BEPCII will be soon delivered     
to the Chinese funding agency. Part of the R\&D work for both machine and
detector has been going for half a year. The expected time for
the physics run with BESIII is around the year of 2005-2006.   

\begin{figure*}[htbp]
\centerline{
\epsfig{file=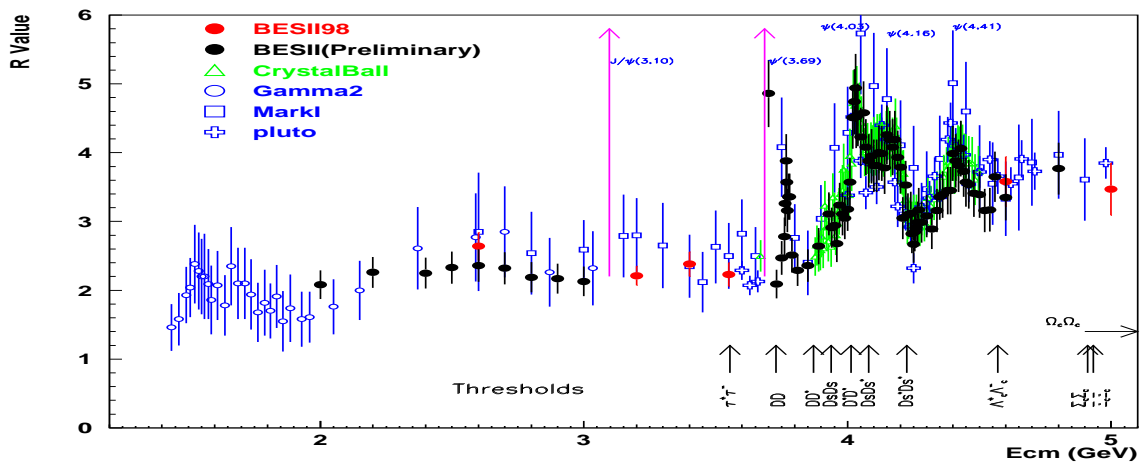,height=6cm,width=15cm}}
\caption{Physics future in 2-5 GeV.}
\label{fig:r2t5}
\end{figure*}

BEPCII, together with CESR for CLEOC~\cite{cleoc}, will be tau-charm 
factory machines for tau-charm physics, including the testing of 
QCD in the energy region
between 2-5 Gev. The physics features in this energy region, as shown in 
Figure~\ref{fig:r2t5}, are significant
and striking, mainly: (1) many resonances and particle
pairs can be directly produced at their thresholds; (2) it's a region of
transition
between smooth and resonance structure, between pQCD and QCD; (3) it's a region where 
gluonic matter, glueballs, hybrids and exotic states are considered to be
located.  

BESIII and CLEOC can collect data as shown in
Table~\ref{tab:cleoc_bes3}. 

\begin{table*}[hbtp]
\caption{Data samples collected with BESI and BESII, and maybe 
collected with CLEOC and BESIII.}
\begin{center}
\begin{tabular}{|cccc|} \hline
$E_{cm}$(GeV) & Physics & BESI+BESII & CLEOC and BESIII\\
\hline
3.1   & $J/\psi$    & $7.8 \times 10^6 + 5 \times 10^7$ & $10^9-10^{10}$ \\
3.55  & $\tau$       & $5~pb^{-1}$       &  $>10^6$ \\ 
3.69  & $\psi(2S)$   & $3.9 \times 10^6$ &  $10^8-10^9$\\
3.77  & $\psi(3770)$ &                 &  $10^7$ \\
4.03, 4.14 & $\tau,~D_{s}^{+}\bar{D_s^-},~D^0\bar{D^{*0}}$ & $22.3~pb^{-1}$ 
& $10^6$\\
4.6   & $\Lambda_c \bar{\Lambda_c}$ & $10^5-10^6$ \\
2-5   & $R$ scan     & 6+85 points (6.6\%) & 2-3\% \\\hline
\end{tabular}
\end{center}
\label{tab:cleoc_bes3}
\end{table*}

These high statistics data samples are the best laboratory to elucidate 
the tricky situation in light hadron spectroscopy and offer unique 
opportunities for QCD studies and probing possible new physics. The major
physics programs are:

\begin{enumerate}

\item  Systematically study meson spectroscopy, $q\bar{q}$ and 
excited baryonic states, search for $^1P_1$, $\eta_c'$, glueballs, and 
exotic states;

\item  Study the interaction with charmed mesons and baryons, 
measure the absolute branching fraction of D and Ds decays, decay 
constant $f_D,~f_{D_s}$, and CKM element (involving c quark).

\item  Carry out a new study of the $\tau$ lepton. Such as lower the limit
on $m_{\mu_{\tau}}$, determine $m_{\tau}$ to less than 1 MeV, study of
$\tau$ weak current. 

\item Measure the values of $R$ to a precision of 1-3\% level. Test of QCD 
by investigating hadronic event shape and hadron production.           
 
\item Probe possible new physics, such as $D^0\bar{D^0}$ mixing, CP and LFV
processes in $\tau,~J/\psi,~\psi(2S)$ decays, and rare decays (like $J/\psi
\rt DX$, Non-SM $\tau$ decays). 

\section{Summary}

Values of $R$ in 2-5 GeV have been improved to a precision of 6.6\% by the
BES collaboration. A 50 M $J/\psi$ event sample with good quality has been
accumulated with BESII. The data analysis of this data is on going. The
short term future plan for BES collaboration is to run BESII for another
2-3 years. The plan for the long term future is to significantly upgrade
both the detector (BESIII) and machine (BEPCII).  
Physics in the tau-charm energy region is still very rich in B-factory era. Many results need
to be improved, and there are many things to be searched for. Both CLEOC
and BESIII are badly needed to perform the physics in the tau-charm region. 
And both CLEOC and BESIII will be the eminent players and contributors to the
physics in tau-charm energy region in the world in the next 5-10 years.  
Last but not least, it would be extremely important to 
have an $R$ scan measurement from 1-3 GeV with a precision of 
a few percent.  

\section{Acknowledgments}
I would like to thank the organizers of the {\em $e^+
e^-$ Physics at Intermediate Energies Workshop} for their invitation and 
hospitality. I want also to thank my friends and colleagues F.A. Harris,
Xiaoyan Shen, Guangshun Huang for their help on this paper.  

\end{enumerate}

\end{document}